\documentclass{article}

\usepackage{amsfonts}

\usepackage{graphics}

\usepackage{amssymb}

\usepackage{amsmath}

\usepackage{amssymb}

\usepackage{amscd}

\usepackage{euscript}

\usepackage{color}

\newtheorem{Theorem}{Theorem}

\newtheorem{Example}[Theorem]{Example}

\newtheorem{Definition}[Theorem]{Definition}

\newtheorem{Lemma}[Theorem]{Lemma}

\newtheorem{Proposition}[Theorem]{Proposition}

\newtheorem{Fundamental Theorem}{Fundamental Theorem}

\def \A {\mathcal{A}}

\def \lX {\mathfrak{X}}

\def \a {\alpha}

\def \GL {{\rm GL}}

\def \V {\EuScript{V}}

\def \u {\mathfrak{u}}

\def \C {\mathbb{C}}

\def \Der {\mathrm{Der}}

\def \e {\mathfrak{h}}

\def \id {\mathrm{id}}

\def \lg {\mathfrak {g}}

\def \X {\EuScript{X}}

\def \f {\phi}

\def \d {\partial}

\def \a {\alpha}

\def \b {\beta}

\def \g {\gamma}
\def \G {\Gamma}

\def \t {\triangleright}

\def \Gc {\mathcal{G}}

\def \ra {\xrightarrow}

\def \le {\mathfrak{h}}

\def \D {\mathcal{D}}

\def \T { \EuScript{T} }

\def \Tc { \mathcal{T} }

\def \H {\mathcal{H}}

\def \fo {\textrm{ for  each }}

\def \an {\textrm{ and }}

\def \GL {\mathrm{GL}}
\def \gl {\mathfrak{gl}}
\def \F {\mathrm{F}}
\def \G {\EuScript{G}}
\def \H {\EuScript{F}}
\def \an {\textrm{ and }}
\def \Hom {\mathrm{Hom}}
\def\be{\begin{equation}}
\def\ee{\end{equation}}
\def\bea{\begin{eqnarray}}
\def\eea{\end{eqnarray}}
\def \we {\textrm{ where }}
\def \Aut {\mathrm{Aut}}
\def \Ad {\mathrm{Ad}}
\def \h {\mathfrak{h}}

\begin{document}

\title{Lie crossed modules and gauge-invariant actions for {2-BF} theories}

\author{Jo\~{a}o  Faria Martins \footnote{This author was supported  by CMA, through Financiamento Base 2009 ISFL-1-297 from FCT/MCTES/PT.   Work supported by the FCT grants PTDC/MAT/101503/2008  and PTDC/MAT/098770/2008.} \\{ \footnotesize Departamento de Matem\'{a}tica }\\ {\footnotesize 
Faculdade de Ci\^{e}ncias e Tecnologia,   Universidade Nova de Lisboa}\\
{\footnotesize  Quinta da Torre,
2829-516 Caparica,
Portugal } \\
{\it \small jn.martins@fct.unl.pt }
 \\ \\Aleksandar Mikovi\'c \footnote{Member of the Mathematical Physics Group, University of Lisbon. Work supported by the FCT grants PTDC/MAT/099880/2008 and PTDC/MAT/69635/2006.}\\
{\footnotesize {Departamento de Matem\'atica}}\\{\footnotesize {Faculdade de Engenharia e Ci\^{e}ncias Naturais}}\\ {\footnotesize {Universidade Lus\'{o}fona de Humanidades e Tecnologia,}}\\ {\footnotesize {Av do Campo Grande, 376, 1749-024 Lisboa, Portugal}}\\ { \it \small  {amikovic@ulusofona.pt}}}
\maketitle

\begin{abstract}
We generalize the BF theory action to the case of a general Lie crossed module $(\d: H \to G,\t)$, where  $G$ and $H$ are non-abelian Lie groups. Our construction requires the existence of $G$-invariant non-degenerate bilinear forms on the Lie algebras of $G$ and $H$ and we show that there are many examples of such Lie crossed modules by using the construction of crossed modules provided by short {chain complex}es of vector spaces. We also generalize this construction to  {an arbitrary chain complex} of vector spaces, of finite type. We construct two gauge-invariant actions for 2-flat and fake-flat 2-connections with auxiliary fields. The first action is of the same type as the BFCG action introduced by Girelli, Pfeiffer and Popescu for a special class of Lie crossed modules, where $H$ is abelian. The second action is an extended BFCG action which contains an additional auxiliary field. However, these two actions are related by a field redefinition. We also construct a three-parameter deformation of the extended BFCG action, which we believe to be relevant for the construction of non-trivial invariants of knotted surfaces embedded in the four-sphere.
\end{abstract}

\maketitle

\section{Introduction}
Crossed modules, or equivalently 2-groups, have become recently an object of intense study in the context of higher gauge theory, since 2-groups offer a natural way to generalize the physics and the geometry of ordinary gauge theories, see \cite{BaH}. {The corresponding concept of a 2-bundle with {a Lie crossed module as the fiber,  or equivalently, the concept of  an  abelian or non-abelian gerbe with a connection}, was studied in \cite{ACG,BS1,BrMe,Hi,MP}. It was observed in  \cite{BS1,BS2,GP} that {the} vanishing of the fake curvature 2-tensor permits a construction of  a  non-abelian surface holonomy and a precise construction of the surface holonomy was realized in \cite{SW1,SW2,FMP1,FMP2}. It was also observed in \cite{BaH} that the holonomy of a 2-connection with a non-zero fake curvature tensor can be addressed in the context of 3-connections and the corresponding 3-dimensional holonomy; \cite{FMP3}. All these objects can be understood  within the framework of higher category theory.

Given a Lie crossed module $(\d\colon H \to G,\t)$ where $\t$ is a smooth action of the Lie group $G$ on the Lie group $H$ by automorphisms, see {\cite{B1,BC,BL,FMP1}}, an associated differential crossed module $(\d \colon \h \to \lg,\t)$ can be constructed, where $\t$ is now a left action of the Lie algebra $\lg$ of $G$ on the {underlying vector space of the Lie algebra $\h$ of $H$ by derivations}. On a manifold $M$, a (local) 2-connection is given by a $\lg$-valued 1-form $A \in \A^1(M,\lg)$ and an  $\h$-valued 2-form $\b\in \A^2(M,\le)$. The corresponding fake curvature {and 2-curvature tensors} can be written as
\begin{align*}
\H_{(A,\b)}&={\rm F}_A -\d(\b) ,\\
\Gc_{(A,\b)}&=d \b +A \wedge^\t \b
\end{align*}
where ${\rm F_A}=dA+A \wedge A$ is the curvature of $A$ (for  notation and  conventions see subsection \ref{pre}). {A local 2-connection $(A,\b)$ will be called fake-flat if the fake curvature tensor vanishes. Similarly, $(A,\b)$ is called 2-flat if the 2-curvature tensor vanishes.}

 We will be only interested in the local aspects of 2-connections, while the interested reader can consult \cite{ACG,BS1,BrMe,FMP2} for the global properties.  For the case of abelian gerbes and their holonomy see \cite{MP,Hi}. 
 
We will consider two types of gauge transformations for a (local) 2-connection  $({{A}},\b)$.  These transformations  appear in the context of 2-connections on 2-bundles when we pass from one coordinate neighborhood to another. 
Given a smooth map $\f\colon M \to G$, let 
 \begin{equation} {{A}} \mapsto \f^{-1} {{A}}\f+\f^{-1} d \f \,,\quad \b\mapsto \f^{-1} \t \b \,.\label{tgt}\end{equation}
 Given a 1-form $\eta\in \mathcal{A}^1(M,\le)$, let
\begin{equation} {{A}} \mapsto {{A}}+\d\eta\,,\quad\b\mapsto \b +d \eta + {{A}} \wedge^\t \eta+\eta\wedge \eta \, .\label{fgt}\end{equation}
We will call the  transformations  (\ref{tgt}) and  (\ref{fgt}) thin and fat, respectively.

The group of gauge transformations will be given by all pairs $(\f,\eta)$ and the group product will be a semidirect product 
$$(\f,\eta)(\f',\eta')=(\f\f',\f \t \eta'+\eta) \,.$$ 
The thin and the fat gauge transformations provide a right action of the gauge group on the set of local 2-connections.

The action of the thin gauge transformations on the curvature $\F_A$, the 2-curvature $\G_{(A,\b)}$ and the fake curvature $\H_{(A,\b)}$ is given by 
$$\F_{{A}} \mapsto \f^{-1} \F_{{A}}\f\, \quad \G_{({{A}},\b)} \mapsto \f^{-1} \t \G_{({{A}},\b)} \,,\quad \H_{({{A}},\b)} \mapsto \f^{-1} \H_{({{A}},\b)}\f\,.$$
The action of the fat gauge transformations on these fields is not straightforward to find, see \cite{BrMe,ACG,BS2} {and below. It} is given by 
$$\F_{{A}} \mapsto \F_{{A}} + \d \left(d\eta+ {{A}} \wedge^\t \eta  + \eta \wedge\eta \right)\,,$$
$$ \H_{({{A}},\b)}  \mapsto \H_{({{A}},\b)} \,,\quad  
\G_{({A},\b)} \mapsto \G_{({A},\b)}+\H_{({{A}},\b)} \wedge^\t \eta\,.$$

Constructing an action for a theory of {fake-flat and 2-flat} 2-connections is important for the quantization of the theory and consequently for constructing new manifold invariants, as well as for quantizing gravity, see \cite{GPP,BaH}. In \cite{GPP} a gauge-invariant action was constructed for a special class of crossed modules $(\d\colon H \to G,\t)$, such that $H$ is the abelian group associated to the vector space $\lg$ and $\t$ is the adjoint action of $G$ on $\lg$. {Also $\d(X)=1_G$ for each $X \in \lg$.} {In the associated differential crossed module $(\d \colon \h \to \lg,\t)$, $\h$ is an abelian Lie algebra which is given by the underlying vector space of $\lg$, with trivial bracket}. The corresponding action can be written as
\begin{equation}\label{BFCG}
S_0 =\int_{M} \langle B \wedge \H_{(A,\b)} \rangle_\lg +  \langle C \wedge \G_{(A,\b)} \rangle_\lg \,,
\end{equation}
where $B \in \A^2(M,\lg)$, $C\in \A^1(M,\le)$ and $\langle,\rangle_\lg$ is a $G$-invariant non-degenerate bilinear form  in $\lg$. 

The action $S_0$ defines the dynamics of a theory of {2-flat} and fake-flat 2-connections, and we will refer to this type of actions as a BFCG action. $S_0$ is invariant under the thin gauge transformations (\ref{tgt})  if 
\begin{equation} B \mapsto \f^{-1} B \f \,,\quad C \mapsto \f^{-1} \t C \,,\label{bct}\end{equation} 
while the invariance under the fat gauge transformations (\ref{fgt}) requires
\begin{equation}B \mapsto B - [C,\eta] \,,\quad C \mapsto C \,.\label{sbcf}\end{equation}

In this article we will generalize the BFCG action $S_0$ to the case of a general Lie crossed module, i.e. the case when the group $H$ is non-abelian {and the morphism $\d\colon H \to G$ is non-trivial}. Our construction requires the existence of a non-degenerate bilinear form on $\h$ such that it is $G$-invariant. We will show that there are many examples of such Lie crossed modules by using the construction of crossed modules provided by short {chain complex}es of vector spaces. {We will also extend this construction to the case of {an arbitrary} finite type {chain complex} of vector spaces.}

We will construct two gauge invariant actions for 2-flat and fake-flat 2-connections with auxiliary fields. One of them will be a BFCG action, and it will require a generalization of the fat gauge transformations for the $B$ field found in \cite{GPP}, see (\ref{sbcf}). The second action, which will be called extended BFCG action, will require an additional auxiliary field. However, we will show that these two actions are related by a field redefinition. 

We also construct a three-parameter deformation of the extended BFCG action, which we believe to be relevant for the construction of non-trivial invariants of knotted surfaces embedded in the four-sphere.

\section{Lie crossed modules}

In this section we are going to give the necessary definitions and the properties of Lie crossed modules which will be needed for the construction of a generalized BFCG action.
All Lie groups and Lie algebras considered here will be finite-dimensional. For a more detailed exposition of Lie crossed modules see \cite{B1,FMP1,BL}. {For general facts about crossed modules see \cite{BHS,FM1}.}

\begin{Definition}[Lie crossed module]\label{LCM}
 A crossed module ${\X= ( \d\colon H \to  G,\t)}$ is given by a group morphism $\d\colon H \to G$ together with a left action $\t$ of $G$ on $H$ by automorphisms, {such that}:
\begin{enumerate}
 \item $\d(g \t h)=g \d(h)g^{-1}; \fo g \in G \an  h \in H,$
  \item $\d(h) \t h'=hh'h^{-1};\fo  h,h' \in H.$
\end{enumerate}
If both $G$ and $H$ are Lie groups, $\d\colon H \to G$ is a smooth morphism,  and the left action of $G$ on $H$ is smooth then $\X$ will be called a Lie crossed module.
\end{Definition}
We will call $G$ the base group of $\X$ and we will be mainly interested in the case when the base group $G$ is compact in the real case, or has a compact real form in the complex case. This ensures that $G$-invariant non-degenerate bilinear forms in $\lg$ and $\h$ {always} exist; see Definition \ref{invariant} and Lemma \ref{compact}.

A morphism $\X \to \X'$ between the crossed modules ${\X= ( \d\colon H \to  G,\t)}$  and {$\X'=(\d'\colon H' \to G',\t')$} is given by a pair of maps $\f\colon G \to G'$ and  $\psi\colon H \to H'$ such that the diagram
$$ \begin{CD}
  H @>\d>> G \\
 @V \psi VV   @VV \f V \\
  H' @>\d'>> G' \\
   \end{CD}
$$
is commutative. In addition, we must have $\psi(g \t h)=\f(g) \t' \psi(h)$ for each $ h \in H$ and each $ g \in G$.

\begin{Example}
 Let $G$ be a Lie group and $V$ a vector space carrying a  representation $\rho$ of $G$. Then
$(V \ra{v \mapsto 1_G} G,\rho)$ is a crossed module. 
\end{Example}
\begin{Example}
 Let $G$ be a connected Lie group and $\Aut(G)$ the automorphisms Lie group of $G$. The group $\Aut(G)$ has a left action in  $G$ by automorphisms $f \t g=f(g)$, where $f \in \Aut(G)$ and $g \in G$. Together with the  map $g \in G \mapsto \Ad_g$ which sends $g \in G$ to the automorphism $h \mapsto ghg^{-1}$ this defines a crossed module. 
\end{Example}

\subsection{Crossed modules of Lie algebras}
Given a Lie crossed module ${\X= ( \d\colon H \to  G,\t)}$, then the induced Lie algebra map $\d\colon \e \to \lg$, together with the derived action of $\lg$ on $\e$ (also denoted by $\t$) is a differential crossed module, in the sense of the following definition, see \cite{BS1,BS2,B1,FMP1,FMP2,BC}. 

\begin{Definition}[Differential crossed module] A differential crossed module ${\mathfrak{X}=(\d \colon \e \to  \lg,\t )}$ is given by a Lie algebra morphism $\d\colon \e \to \lg$ together with a left action of $\lg$ on the underlying vector space of  $\e$, {such that}:
\begin{enumerate}
 \item For any $X \in \lg$ the map $\xi \in \e \mapsto X \t \xi \in \e$ is a derivation of $\e$, which can be written as 
$$X \t [\xi,\nu]=[X \t \xi,\nu]+[\xi, X \t \nu]\,,\quad \forall\, X \in \lg\,, \forall\, \xi ,\nu\in \e \,.$$
\item The map $\lg \to \Der(\e)$ from $\lg$ into the derivation algebra of $\e$ induced by the action of $\lg$ on $\e$ is a Lie algebra morphism, which can be written as
$$[X,Y] \t \xi=X \t (Y \t \xi)-Y \t(X \t \xi)\,,\quad \forall\, X,Y \in \lg \,,\forall\, \xi \in \e \,,$$
\item 
\begin{equation}
\d( X \t \xi)= [X,\d(\xi)]\,,\quad \forall X \in \lg\,, \,\forall\, \xi\in \e\,,\label{fid} \end{equation}
\item \begin{equation} \d(\xi) \t \nu=[\xi,\nu]\,,\quad  \forall\, \xi,\nu\in \e\,.\label{sid}\end{equation}
\end{enumerate}
\end{Definition}
Therefore, given a differential crossed module ${\mathfrak{X}=(\d \colon \e \to  \g,\t )}$, there exists a unique crossed module of simply connected Lie groups ${\X=(\d \colon H \to  G,\t )}$ which corresponds to $\mathfrak{X}$, up to an isomorphism.

A very useful identity satisfied in any differential crossed module is
\begin{equation*}
\d(\xi)\t \nu=[\xi,\nu]=-[\nu,\xi]=-\d(\nu) \t\xi\,,\quad \forall\,  \nu,\xi \in \e.
\end{equation*}
\subsubsection{Mixed relations}\label{mixed}
Let $\X=(\d\colon H \to G,\t)$ be a Lie crossed module, and let $\lX=(\d\colon\h \to \lg,\t)$ be the associated differential crossed module. Therefore $G$ acts on $\lg$ by the adjoint action, and on $\le$ by the action induced by $\t$. The following mixed relations are satisfied
$$\d(g \t \xi)=g\d(\xi)g^{-1}\,,\quad g \in G \,, \xi \in \le \,, $$

$$\d(X \t h)=X \d(h)-\d(h)X \,, \quad X \in \le \,, h \in H \,,$$

$$\d(h) \t \xi=h\xi h^{-1}\,, \quad h \in H \,, \xi \in H \,, $$
and
$$\d(\xi) \t h=\xi h-h\xi\,, \quad \xi \in \le \,, h \in H \,.$$

An identity that will be used to prove the gauge invariance of the extended BFCG action is 
\begin{equation}\label{eqforinvariance}
g \t (X \t \xi)=\left(gXg^{-1}\right) \t\left (g \t \xi\right)\,, \quad \forall g \in G, \xi \in \h, X \in \lg.
\end{equation}

\subsubsection{$G$-invariant bilinear forms}\label{ndg}

We will introduce the following $G$-invariant bilinear forms on the Lie algebras of a Lie crossed module in order to be able to construct a generalized {BFCG} action.

\begin{Definition}[non-degenerate symmetric $G$-invariant forms] \label{invariant} Let us consider a Lie crossed module ${\X= ( \d\colon H \to  G,\t)}$ and let $\lX=( \d\colon \le \to  \lg,\t)$ be the associated differential crossed module. A symmetric non-degenerate  $G$-invariant form in $\lX$ is given by a pair of non-degenerate symmetric  bilinear forms $\langle,\rangle_\lg$ in $\lg$ and $\langle, \rangle_\le$ in $\le$ such that
\begin{enumerate}
 \item $\langle,\rangle_\lg$ is $G$-invariant, i.e. 
$$\langle gXg^{-1},gYg^{-1} \rangle=\langle X, Y \rangle\,, \quad\forall\, g \in G \,,\,\, X,Y \in \lg \,,$$
\item  $\langle,\rangle_\le$ is $G$-invariant, i.e.
$$\langle g\t \xi, g \t \nu \rangle=\langle \xi , \nu \rangle\,, \quad\forall\, g \in G \,, \,\,\xi,\nu \in \le \,. $$
\end{enumerate}
\end{Definition}
Note that $\langle,\rangle_\h$ is necessarily $H$-invariant. This is because
\begin{align*}
 \langle h\xi h^{-1},h\nu h^{-1} \rangle_\h=\langle \d(h) \t \xi ,\d(h)\t \nu \rangle_\h=\langle \xi ,\nu \rangle_\h\,, \quad  \forall \xi,\nu \in \le \,,\, h \in H \,,
\end{align*}
where we have used the mixed relations  \ref{mixed}.

There are no compatibility conditions between the  symmetric bilinear forms $\langle,\rangle_\lg$ and $\langle,\rangle_\le$. From the well known fact that any representation of $G$ can be made unitary if $G$ is a compact group, it follows that
\begin{Lemma}\label{compact}
 Let $\X= ( \d\colon H \to  G,\t)$ be a Lie crossed module with the base group  $G$ being compact in the real case, or  having a compact real form in the complex case. Then one can construct $G$-invariant  symmetric non-degenerate bilinear forms $\langle,\rangle_\lg$ and $\langle, \rangle_\le$ in the associated differential crossed module $\mathfrak{X}=( \d\colon \le \to  \lg,\t)$. Furthermore these forms can be chosen to be positive definite.
\end{Lemma}

Given  $G$-invariant {symmetric} non-degenerate bilinear forms in $\lg$ and $\le$, we can define a bilinear antisymmetric map $\Tc\colon \le \times \le \to \lg$ by the rule
$$\langle  \Tc(u,v), Z \rangle_\lg=-\langle  u, Z \t  v\rangle_\le\,, \quad u,v \in \le \,,\, Z \in \lg \,.$$
Note that $\langle,\rangle_\lg$ is non-degenerate and 
$$\langle  u, Z \t v\rangle_\le=- \langle Z \t u,  v\rangle_\le =- \langle   v, Z \t u\rangle_\le \,.$$
Morever, given $g \in G$ and $u,v \in \le$, we have:
$$\T(g \t u,g \t v)=g \T(u,v)g^{-1}, $$
since for each $X \in \lg$ and $u,v \in \h$ we have: 
\begin{align*}
\langle X,g^{-1} \T( g \t u,g \t v) g \rangle_\lg&=\langle gXg^{-1},\T(g \t u,g \t v)\rangle_\lg =-\langle (gXg^{-1}) \t g \t  u, g \t v\rangle_\h\\
                                      &=-\langle X \t u,v \rangle_\h  
                                      =\langle X, \T(u,v)\rangle_\lg       
\end{align*}
We thus have the following identity:
$$\T(X \t u,v)+\T(u,X \t v)=[X,\T(u,v)]. $$

The map $\Tc$ will play a major role in the construction  of the generalized BFCG action {and the corresponding gauge transformations.}
  
\subsection{Crossed modules from  {chain complex}es of vector spaces}

We will now show that there exists a rich class of examples of Lie crossed modules $(\partial : H \to G,\t)$ with the desired properties, i.e. $G$ being compact and $H$ a non-abelian group, by constructing them from  {chain complex}es of vector spaces.

\subsubsection{Crossed modules from short {chain complex}es of vector spaces}
The definition of a Lie crossed module from a  short {chain complex} of vector spaces was given in \cite{BC,BL}. We are going to use this definition in order to explicitly construct a Lie crossed module with the desired properties.

{A minor modification of the definition  will yield  a Lie crossed module from {an arbitrary chain complex} of vector spaces; see below.}

Let $\V=(V \ra{{\bf \phi}} U)$ be a short {chain complex} of finite-dimensional vector spaces. This means that $V$ and $U$ are vector spaces and ${\bf \phi}\colon V \to U$ is a linear map.
We can define a crossed module of Lie groups 
$$\GL(\V)=({\d}\colon \GL_1(\V) \to \GL_0(\V),\t) $$  
in the following way:
 
Let $\Hom_0(\V)$ be the algebra of chain maps $f\colon \V \to\V$, such that the composition of maps is the algebra product. A chain map $f\colon \V \to \V$ is defined by a pair of linear maps $(f_V,f_U)$, where $f_U \colon U \to U$ and $f_V\colon V \to V$, such that the diagram 
 $$\begin{CD} V @>{\bf \phi}>> U\\
              @V {f_V} VV  @VV {f_U} V\\
    V @>>{\bf \phi}> U
  \end{CD}
 $$
is commutative. This is equivalent to ${\bf \phi}\circ f_V= f_U \circ {\bf \phi}.$ Note that if $F=(f_V,f_U)$ and  $F'=(f_V',f_U')$ are chain maps, then their composition  $F \circ F'= (f_V\circ f_{V'},f_U\circ f_{U'}) $ is also a chain map.

Consider the set  $\GL_0(\V)$ of   invertible elements of $\Hom_0(\V)$, which is the set of pairs $F=(f_V,f_U)$ such that $f_U\colon U \to U$ and $f_V\colon V \to V$ are invertible linear maps. Note that this is an open subset of $ \Hom_0(\V)$.
Then  $\GL_0(\V)$  is a Lie group under the  composition of chain  maps. More precisely, $\GL_0(\V)$  is a closed subgroup of the general linear group $\GL(V\oplus U)$. The  Lie algebra  $\gl_0(\V)$ of $\GL_0(\V)$ is identical to $\Hom_0(\V)$ as a vector space,  with the bracket given by the associative algebra structure of $\Hom_0(\V)$, so that $[F,F']=F\circ F'-F' \circ F$, for chain maps $F,F'\colon \V \to \V$.

Consider the semigroup $\Hom_1(\V)$ of all linear maps $s\colon U \to V$, with a product defined as 
$$s*t=s+t+s{\bf \phi} t \,.$$
This is an associative product, with the unit being the null linear map $U \to V$. The map ${\d}\colon \Hom_1(\V) \to \Hom_0(\V)$ such that
$${\d}(s)=(s {\bf \phi}+\id,{\bf \phi} s +\id), \we s \in \Hom_1(\V)$$
preserves the products. The set $\GL_1(\V)$  of linear maps $s\colon U \to V$ for which ${\d}(s)$ is invertible is certainly open in $\Hom_1(\V)$, since ${\d}\colon \Hom_1(V) \to \Hom_0(\V)$ is continuous, and 
$\GL_0(\V)$ is open in  $\Hom_1(\V)$. Furthermore, since ${\d}$ preserves the products, if $s$ and $t$ are in $\GL_1(\V)$, then so is $s*t$. If $s\in \GL_1(\V)$, then the inverse $\overline{s}$ of $s$ with respect to $*$, whose unit is the null map $U \to V$, is
$$\overline{s}=-(\id+s {\bf \phi})^{-1}s=-s (\id +{\bf \phi} s)^{-1}\, .$$

Notice that ${\d}(\overline{s})$ is invertible if ${\d}(s)$ is invertible.
Therefore $\GL_1(\V)$ is a Lie group of dimension $\dim(U)\times \dim(V)$, and ${\d}\colon \Hom_1(\V)\to \Hom_0(\V)$ is a Lie group morphism.

The Lie algebra $\gl_1(\V)$ of $\GL_1(\V)$ is given by the vector space $\Hom_1(\V)$ of all maps $s\colon U \to V$, with the bracket given by 
$$[s,t]=s{\bf \phi} t-t{\bf \phi} s \,.$$ 
The map ${\d}\colon \gl_1(\V) \to \gl_0(\V)$ such that
$${\d}(s)=(s {\bf \phi},{\bf \phi} s) $$ 
is a morphism of Lie algebras, and it is exactly the derivative of ${\d}$.

A left action of {$\GL_0(\V)$ on $\GL_1(\V)$,} by automorphism can be defined as
$$(f_V,f_U) \t s=f_V sf_U^{-1} \,$$ 
where $s \in \Hom_1(\V)$ and $(f_U,f_V) \colon \V \to \V$ is an invertible chain map. Therefore if  $F=(f_V,f_U)$ is invertible and ${\d}(s)$ is invertible, then so is: $${\d}(F \t s)=\left(f_V {\d}(s) f_V^{-1}, f_U {\d}(s) f_U^{-1}\right)=F {\d}(s) F^{-1} \,.$$ 
The differential form of this action is the left action of {$\gl_0(\V)$ on $\gl_1(\V)$} by derivations such that:
$$(f_V,f_U) \t s=f_V s- sf_U \,.$$ 

To finish proving this construction defines a crossed module note that if $s,t \in {\GL_1(\V)}$
\begin{align*}
 {\d}(s) \t t =(s {\bf \phi} +\id) t ({\bf \phi} s +\id)^{-1} \,,
\end{align*}
whereas:
\begin{align*}
 (s*t*\overline{s})&=(s +t+s {\bf \phi} t)*\overline{s}=s+\overline{s} + s {\bf \phi} \overline{s}+t+s {\bf \phi} t + t {\bf \phi} \overline{s}+s{\bf \phi} t {\bf \phi} \overline{s}\\
&=t+s {\bf \phi} t + t {\bf \phi} \overline{s}+s{\bf \phi} t {\bf \phi} \overline{s} \,,
\end{align*}
and therefore, since $\overline{s}=-s (\id +{\bf \phi} s)^{-1}$:
\begin{align*}
 (s*t*\overline{s}) ({\bf \phi} s +\id)=t {\bf \phi} s +t +s{\bf \phi} t{\bf \phi} s +s {\bf \phi} t-t {\bf \phi} s - s {\bf \phi} t {\bf \phi} s=(s {\bf \phi} +\id) t \,.
\end{align*}

We have thus given a vector space map $\V=(V \ra{{\bf \phi}} U)$, a short {chain complex} of vector spaces, defined a Lie crossed module 
$$\GL(\V)=\left({\d}\colon \GL_1(\V) \to \GL_0(\V),\t\right)\,, $$
whose differential form is the differential crossed module
$$\gl(\V)=\left({\d}\colon \gl_1(\V) \to \gl_0(\V),\t\right)\,. $$
{The proof that $\gl(\V)$ is a differential crossed module is completely analogous to the case of the crossed module.}

\begin{Example}
 Consider the case when $V=U=\C^2$ and ${\bf \phi}=\id$. Then $\GL_0(\V)=\GL_1(\V)=\GL(\C^2)$ and ${\d}$ is the identity map.
\end{Example}

\begin{Example}
 Consider the case when $V=U=\C^2$ and ${\bf \phi}=0$. Then $\GL_0(\V)=\GL(\C^2)\times \GL(\C^2)$. On the other hand $\GL_1(\V)=\Hom(\C^2,\C^2)$, the space of all linear maps $\C^2 \to \C^2$, with left action $(A,B) \t f= A f B^{-1}$. In this case  ${\d}(f)=1$ for each $f \in \GL_1(\V)$, and the (abelian) group structure in $\GL_1(\V)$ is the usual sum of linear maps.
\end{Example}
\begin{Example}
 Let $X,Y,Z$ be vector spaces, The most general case of the crossed module defined by a short {chain complex} of vector spaces is given by the case when $V=X\oplus Y$ and $U=X\oplus Z$, and ${\bf \phi}(x,y)=(x,0)$. A map $V \to V$ is given by a matrix $\begin{pmatrix} A\colon X \to X & B\colon Y \to X\\C\colon X \to Y & D\colon Y \to Y\end{pmatrix}$ of linear maps, and analogously for a map $U \to U$, which is given by a matrix $\begin{pmatrix} A\colon X \to X & B\colon Z \to X\\C\colon X \to Z & D\colon Z \to Z\end{pmatrix}$ of linear maps. The Lie algebra $\gl_0(\V)$ is given by all pairs of linear maps of the form, $$(f_U,f_V)=\left  ( \begin{pmatrix} A & 0\\C & D\end{pmatrix},\begin{pmatrix} A & B'\\0 & D'\end{pmatrix}  \right), $$
with commutator being the usual commutator of matrices.
\begin{multline*}
\left [ \left  ( \begin{pmatrix} A_1 & 0\\C_1 & D_1\end{pmatrix},\begin{pmatrix} A_1 & B_1'\\0 & D'_1\end{pmatrix}  \right) , \left  ( \begin{pmatrix} A_2 & 0\\C_2 & D_2\end{pmatrix},\begin{pmatrix} A_2 & B'_2\\0 & D'_2\end{pmatrix}  \right) \right] 
\\=\left ( \left [ \begin{pmatrix} A_1 & 0\\C_1 & D_1\end{pmatrix}  ,  \begin{pmatrix} A_2 & 0\\C_2 & D_2\end{pmatrix}  \right ] , \left [   \begin{pmatrix} A_1 & B_1'\\0 & D'_1\end{pmatrix}   ,    \begin{pmatrix} A_2 & B'_2\\0 & D'_2\end{pmatrix}   \right] \right ).
 \end{multline*}
The Lie algebra $\gl_1(\V)$ is given by all matrices of linear maps of the form:
$$s=\begin{pmatrix} A\colon X \to X & B\colon Z \to X\\C\colon X \to Y & D\colon Z \to Y\end{pmatrix},$$
with boundary $$\b'(s)=\left ( \begin{pmatrix} A & 0\\ C & 0 \end{pmatrix}, \begin{pmatrix} A & B\\  0 & 0 \end{pmatrix}\right)$$
and commutator (not coinciding with the commutator of matrices):
$$\left [   \begin{pmatrix} A & B \\C  & D\end{pmatrix} ,\begin{pmatrix} A' & B' \\C'  & D'\end{pmatrix} \right]=\begin{pmatrix} [A,A'] &  AB'-A'B \\ CA'-C'A & CB'-C'B\end{pmatrix} .$$
The left action of $\gl_0(\V)$ on $\gl_1(\V)$ by derivations is:
$$ \left  ( \begin{pmatrix} A & 0\\C & D\end{pmatrix},\begin{pmatrix} A & B'\\0 & D'\end{pmatrix}  \right) \t  \begin{pmatrix} A'& B'\\C'& D'\end{pmatrix}= \begin{pmatrix} A & 0\\C & D\end{pmatrix} \begin{pmatrix} A'& B'\\C'& D'\end{pmatrix} -\begin{pmatrix} A'& B'\\C'& D'\end{pmatrix}\begin{pmatrix} A & B'\\0 & D'\end{pmatrix}.$$
It is an easy calculation to prove that this indeed yields a crossed module of Lie algebras.
\end{Example}
In the previous example note that the kernel $\ker {\d}$  of ${\d}\colon \gl_1(\V) \to \gl_0(\V)$ is the vector space of all linear maps $Z \to Y$, with trivial commutator.  The cokernel of ${\d}$ is given by all the pairs of maps $(D \colon Y \to Y,D'\colon Z \to Z)$ with pairwise commutator:
$$[(Z_1,Z'_1),(Z_2,Z_2')]=\left ([Z_1,Z_2],[Z_1',Z_2']\right).$$ The action of $\gl_0$ on $\gl_1$ descends to an action on ${\rm coker}({\bf \phi})$ on  $\ker {\bf \phi}$ which has the form:
$$(Z,Z') \t D=ZD-DZ' .$$

\subsubsection{The compact case}\label{cc}
Given a linear map of vector spaces $\V=(V \ra{{\bf \phi}} U)$, the group  $\GL_0(\V)$ is clearly non-compact. Therefore the lemma \ref{compact} cannot be applied to the crossed module $\GL(\V)=\left({\d}\colon \GL_1(\V) \to \GL_0(\V),\t\right)$. However we can easily modify the construction of the crossed module $\GL(\V)$ in order to get a compact base group.
\begin{Definition}\label{bforms}
Given a map $\V=(V \ra{{\bf \phi}} U)$ of vector spaces, an inner product for $\V$ is simply given by two non-degenerate positive definite  bilinear forms $\langle,\rangle_U$ and $\langle,\rangle_V$ in $U$ and $V$, with no further compatibility conditions.
\end{Definition}

Given $\big(\V,\langle,\rangle_V,\langle,\rangle_U\big)$ as above, define a Lie group ${\rm U}_0\big(\V,\langle,\rangle_V,\langle,\rangle_U\big)$ as being given by all pairs $F=\big(f_V,f_U\big) \in \GL_0(\V)$ for which $f_V\colon V \to V$ and $f_U\colon U \to U$ each are unitary. Clearly ${\rm U}_0\big(\V,\langle,\rangle_V,\langle,\rangle_U\big)$ is a closed (in fact compact) Lie subgroup of $\GL_0(\V)$. In addition, since ${\d}\colon \GL_1(\V) \to \GL_0(\V)$ is a Lie group morphism, $${\rm U}_1\big(\V,\langle,\rangle_V,\langle,\rangle_U\big)
\doteq{\d}^{-1}\Big( {\rm U}_0\big(\V,\langle,\rangle_V,\langle,\rangle_U\big)\Big)$$
is a Lie subgroup of $\GL_1(\V)$.

Given $s\in {\rm U}_1\big(\V,\langle,\rangle_V,\langle,\rangle_U\big)$ and $F=\big(f_V,f_U\big)\in {\rm U}_0\big(\V,\langle,\rangle_V,\langle,\rangle_U\big)$ let us see that $F\t s$ is still in ${\rm U}_1\big(\V,\langle,\rangle_V,\langle,\rangle_U\big)$;
the pair ${\d}(F \t s)$ satisfies:
\begin{align*}
 {\d}(F \t s)&=\big(f_V s f_U^{-1}  {\bf \phi} +\id, {\bf \phi} f_V s f_U^{-1} +\id\big)\\
&=\big(f_V s  {\bf \phi} f_V^{-1} +\id, f_U {\bf \phi} s f_U^{-1} +\id\big)\\
&=\big(f_V (s  {\bf \phi} +\id) f_V^{-1} , f_U ({\bf \phi} s +\id) f_U^{-1} \big)
\end{align*}
thus  $F\t s \in {\rm U}_1\big(\V,\langle,\rangle_V,\langle,\rangle_U\big).$

We have therefore defined a Lie crossed module with compact base group 
$$U(\V,\langle,\rangle_U,\langle,\rangle_V)=\left({\rm U}_1(\V,\langle,\rangle_V,\langle,\rangle_U) \ra{{\d}} {\rm U}_0(\V,\langle,\rangle_V,\langle,\rangle_U),\t\right) $$
given a $(\V,\langle,\rangle_V,,\langle,\rangle_U)$ as in definition \ref{bforms}. The  differential form of this is given by the differential crossed module
$$\u(\V,\langle,\rangle_V,\langle,\rangle_U)=\left(\u_1(\V,\langle,\rangle_V,\langle,\rangle_U) \ra{{\d}} \u_0(\V,\langle,\rangle_V,\langle,\rangle_U),\t\right), $$
defined in a completely analogous way. Therefore $\u_0(\V,\langle,\rangle_V,\langle,\rangle_U)$ is given by all pairs $F=(f_V,f_U)$ of linear maps such that  $f_U+f^*_U=0$, $f_V+f^*_V=0$, and also ${\bf \phi} f_V=f_U {\bf \phi}$.  The commutators are given by  $$[(f_V,f_U),(f_V',f_U')]=([f_V,f_V'],[f_U,f_U']).$$ 
On the other hand $\u_1(\V,\langle,\rangle_V,\langle,\rangle_U)$ is given by all linear maps $s\colon U \to V$ such that $({\bf \phi} s)^*+{\bf \phi} s=0$ and also  $(s{\bf \phi})^*+s{\bf \phi}=0$, {with commutator $[s,t]=s \phi t-t\phi s$}. The left action of $\u_0$ on $\u_1$ by derivations is, as before
$$(f_V,f_U) \t s=f_V s- sf_U.$$ 
Thus if $s \in \u_1$ and $(f_V,f_U)\in \u_0$ then $(f_V,f_U) \t s\in \u_1$.

\begin{Example}
 If $U=V=\C^2$ with standard inner product, and ${\bf \phi}\colon V \to U$ is the identity map then ${\rm U}_0(\V)$ and ${\rm U}_1(\V)$ each are equal to $U(\C^2)$ and ${\d}\colon {\rm U}_1 \to {\rm U}_0$ is the identity map. The action of ${\rm U}_0$ on ${\rm U}_1$ is the action by conjugation.
\end{Example}

\begin{Example}
Consider a vector space $W$ with a positive-definite inner product. Let $U=V=W \oplus W$, and let the map ${\bf \phi}\colon U \to V$ be such that
${\bf \phi}(w,w')=(w,0)$. Then $\u_0(\V)=\u(W) \oplus \u(W) \oplus \u(W)$ and
$\u_1(\V)= \u(W)\oplus\gl(W)$, with ${\d}(w_1,w_2,w_3)=(w_1,0,0)$. 
\end{Example}
{In the previous example, to get genuinely new crossed modules, we should consider the case when $U$ and $V$ have orthogonal decompositions
$U=X\oplus Y$ and $V=X' \oplus Z$, and where $\f(x,y)=(x,0)$. Here $X'$ is $X$ as a vector space, but with a different inner product.}

\subsubsection{{The differential 2-crossed module given by  {an arbitrary chain complex} of vector spaces}}\label{gc}
{Any {chain complex} of finite-dimensional vector spaces} $${\V=(\ldots \ra{{\bf \phi}}  V_{n+1} \ra{{\bf \phi}} V_n \ra{{\bf \phi}} V_{n-1} \ra{{\bf \phi}}    \ldots),}$${with arbitrary, albeit finite, length, also} gives a differential crossed module $$\gl(\V)=\left ({\d}\colon \gl_1(\V) \to \gl_0(\V) ,\t \right), $$ thus a crossed module of Lie groups $\GL(\V)$. In the case of 2-crossed modules  this construction appeared in \cite{KP}, for more details see \cite{FMP3}.

The Lie algebra $\gl_0(V)$ is given by all chain maps $f\colon \V \to \V$, with the usual commutator.
A degree $n$ map $h\colon \V \to \V$ is given by a sequence of linear maps $h_i\colon V_i \to V_{i+n}$, without any compatibility relations with the boundary maps ${\bf \phi}$. Two degree 1-maps $s,t\colon \V \to \V$ (homotopies) are {(2-fold)} homotopic if there exists a degree-2 map $h\colon \V \to \V$ {(a 2-fold homotopy)} such that $$s_i(v)-t_i(v)={\bf \phi} h (v)-h {\bf \phi}(v), \fo v \in V_i.$$

The Lie algebra $\gl_1(\V)$ is defined as being   the vector space of  all degree 1 maps $s\colon \V \to \V$  up to  {(2-fold)}  homotopy. The boundary map ${\d}\colon \gl_1(\V) \to \gl_0(\V)$ is
$${\d}(s)=s{\bf \phi}+{\bf \phi} s ,$$
and the bracket is $$[s,t]=s{\bf \phi} t -t {\bf \phi} s +st{\bf \phi}-ts {\bf \phi}.$$
(This is antisymmetric and satisfies the Jacobi relation on the nose, i.e. before passing to the quotient.) It is easy to see that ${\d}$ is a Lie algebra morphism.

The action of ${\gl}_0(\V)$ on ${\gl}_1(\V)$ is
$$f \t s=fs-sf.$$
This is an action by derivations, before passing to the quotient. We trivially have ${\d}(f \t s)=[f, {\d}(s)]$. Moreover ${\d}(s) \t t-[s,t]={\bf \phi} \circ (st-ts)-(st-ts) \circ {\bf \phi}$, therefore by considering the quotient of the space of degree-1 maps with respect  to {(2-fold)}  homotopy defines a {differential} crossed module $\gl(\V)$.

This construction can be adapted in the obvious way to give a crossed module with a compact base group as in \ref{cc}. {This can be done by picking inner products $\langle, \rangle_n$ is each $V_n$, without any compactibility relations with the boundary maps $\phi$.}

\subsection{{Lie Crossed modules and differential forms}}

In order to perform the calculations more efficiently, we will introduce the component notation.
Let $T_m$ be a basis in $\lg$ and $\tau_\mu$ a basis in $\le$, such that
\be [T_m , T_n ] = f_{mn}^r \,T_r\,,\quad [\tau_\mu , \tau_\nu ] = \f_{\mu\nu}^\rho \,\tau_\rho \,.\ee
Then 
\be \d \tau_\mu = \d_\mu^m \,T_m \,,\quad T_m \t \tau_\mu = \t_{m\mu}^\nu \, \tau_\nu \,,\ee
and the relations (\ref{fid}) and (\ref{sid}) take the following form
\be \t_{m\mu}^\nu \,\d_\nu^r = \d_\mu^n \,f_{mn}^r \,,\label{patr}\ee
\be \d_\mu^m \,\t_{m\nu}^\rho = \f_{\mu\nu}^\rho \,.\label{trpa}\ee

The structure constants satisfy the Jacobi identities
\be f_{mt}^s \,f_{nr}^t = f_{m[n|}^t \,f_{t|r]}^s \,, \quad \f_{\mu\epsilon}^\sigma \,\f_{\nu\rho}^\epsilon = \f_{\mu[\nu|}^\epsilon \,\f_{\epsilon|\rho]}^\sigma \,,\label{jac}\ee
where $X_{[IJ]}= X_{IJ} - X_{JI}$. 

A $\lg$-valued form $X$ on $M$ will be defined as $X=X^m \,T_m$, where $X^m$ are forms on $M$, all of the same degree, and $T_m$ are taken to be the adjoint representation matrices. Let $X$ and $Y$ be two $\lg$-valued forms on $M$. We define
$$X\wedge Y = (X^m \wedge Y^n )\,T_m T_n \,.$$
This matrix will not be in general an adjoint matrix for $\lg$. However,
if $X$ is an odd form, then
$$ X\wedge X = X^m \wedge X^n \,T_m T_n = \frac{1}{2}\,X^m \wedge X^n \,[T_m , T_n] = \frac{1}{2}\,f_{mn}^r \,X^m \wedge X^n \,T_r \in \lg \,,$$
and similarly for an odd form $\xi$ in $\le$. 

Note that $X\wedge X \notin \lg$ if $X$ is an even form and also $X\wedge Y \notin \lg$ for $X\ne Y$. However, if $X$ is a $p$-form and $Y$ a $q$-form, then
$$X\wedge Y + (-1)^l \,Y\wedge X \in \lg \,,$$ 
if $l + pq$ is an odd integer.

If $\xi$ is a $\le$-valued form on $M$ and $X$ a $\lg$-valued form on $M$, then we define
$$ X \wedge^\t \xi = X^m \wedge \xi^\mu \,T_m \t \tau_\mu = X^m \wedge \xi^\mu \,\t_{m\mu}^\nu \tau_\nu \in\le\,.$$ 
One also has
$$ \d\xi = \xi^\mu \d\tau_\mu = \xi^\mu \d_\mu^m \,T_m \,.$$

Let us choose a non-degenerate {bilinear} $G$-invariant form in $\lg$, $\langle,\rangle_\lg$, and a non-degenerate {bilinear} $G$-invariant form  $\langle,\rangle_\le$ in $\le$, see Definition \ref{invariant}. If $X$ and $Y$ are two $\lg$-valued forms on $M$, then we define
$$ \langle X\wedge Y \rangle_\lg = X^m \wedge Y^n \langle T_m , T_n \rangle_\lg = X^m \wedge Y^n Q_{mn}\,.$$
Similarly, for two $\le$-valued forms
$$\langle \xi\wedge \eta \rangle_\le = \xi^\mu \wedge \eta^\nu \langle \tau_\mu , \tau_\nu \rangle_\le = \xi^\mu \wedge \eta^\nu q_{\mu\nu} \,.$$ 
The matrices $Q$ and $q$ correspond to $G$-invariant metrics on  $\lg$ and $\le$, respectively.

As far as the bilinear antisymmetric map $\Tc : \le\times \le \to \lg$  (see \ref{ndg}) is concerned, one can write
$$ \Tc (\tau_\mu , \tau_\nu ) = \Tc_{\mu\nu}^m \, T_m \,,$$
so that the defining relation for $\Tc$ becomes
$$ \Tc_{\mu\nu}^n \, Q_{nm} = \t_{m[\mu|}^\rho \, q_{\rho|\nu]}  \,.$$
Given two $\le$-valued forms $\xi$ and $\eta$, we can define a $\lg$-valued form
$$ \xi \wedge^{\Tc} \eta = \Tc_{\mu\nu}^m \, \xi^\mu \wedge \eta^\nu \,T_m \,.$$ 

\section{Actions for a 2-BF theory}

Our goal now is to construct a gauge invariant action for the theory of {2-flat and fake-flat 2-connections} associated to a general Lie crossed module such that this action is a generalization of a BF theory action \cite{B2}. We will call such a theory a 2-BF theory, since it can be considered as a BF theory for a 2-group.

\subsection{Preliminaries}\label{pre}

{Let us consider a Lie crossed module ${\X= ( \d\colon H \to  G,\t)}$ and let $\lX=( \d\colon \le \to  \lg,\t)$ be the associated differential crossed module.}
Let $M$ be a 4-manifold. Our initial space of fields is given by the forms
\begin{itemize}
 \item ${{A}} \in \mathcal{A}^1(M,\lg)$
 \item $\b \in \mathcal{A}^2(M,\le)$.
\end{itemize}
As we have explained in section 1, the pair $(A,\b)$ will be called a local 2-connection.

Let us introduce the curvature 2-form of ${{A}}$
$$\F_{{A}}=d{{A}} + {{A}} \wedge {{A}} \,, $$
and the 2-curvature 3-form of the pair $({{A}},\b)$
$$\G_{({{A}},\b)}=d \b+ {{A}}\wedge^  \t   \b \,.$$
The corresponding fake curvature tensor is given by
$$\H_{({{A}},\b)}=\F_{{{A}}}-\d\b \,.$$

Given that $(A,\b)$ is a local 2-connections on a 2-bundle, we will consider the following transformations of the pair $({{A}},\b)$, see sect. 1,
\begin{itemize}
 \item {\bf Thin:} For a smooth map $\f\colon M \to G$
 \begin{equation}\label{thin} {{A}} \mapsto \f^{-1} {{A}}\f+\f^{-1} d \f \,,\quad \b\mapsto \f^{-1} \t \b \,.\end{equation}
 \item {\bf Fat:} For a 1-form $\eta\in \mathcal{A}^1(M,\le)$
\begin{equation}\label{fat} {{A}} \mapsto {{A}}+\d\eta\,,\quad\b\mapsto \b +d \eta + {{A}} \wedge^\t \eta+\eta\wedge \eta \, .\end{equation}
\end{itemize}

Under the thin gauge transformations, the curvature, the 2-curvature and the fake curvature change as
$$\F_{{A}} \mapsto \f^{-1} \F_{{A}}\f, \quad \G_{({{A}},\b)} \mapsto \f^{-1} \t \G_{({{A}},\b)} \,,\quad \H_{({{A}},\b)} \mapsto \f^{-1} \H_{({{A}},\b)}\f \,.$$

The action of the  fat gauge transformations on $\F_A$ is given by
$$\F_{{A}} \mapsto \F_{{A}} + d(\d\eta)+ {{A}} \wedge \d\eta + \d\eta\wedge A + \d\eta \wedge \d\eta\,. $$
Since
$$ A\wedge \d\eta + \d\eta\wedge A = \d \left(A\wedge^\t \eta\right) \,,\quad \d\eta\wedge\d\eta = \d(\eta\wedge\eta)\,,$$
due to (\ref{patr}) and (\ref{trpa}), one obtains
$$\F_{{A}} \mapsto \F_{{A}} + \d \left(d\eta+ {{A}} \wedge^\t \eta  + \eta \wedge\eta \right)\,. $$
Therefore  the fake curvature is invariant under the fat gauge transformations
$$\H_{({{A}},\b)} \mapsto \H_{({{A}},\b)} \,. $$

The 2-curvature $\G_{(A,B)}$ transforms under the fat gauge transformations as 
\begin{eqnarray}
 \G &\to& d \b + d (A \wedge^\t \eta) + d (\eta\wedge \eta)\cr
&\,&+\,A \wedge^\t \b +A \wedge^\t d \eta  + A \wedge^\t \left(A \wedge^\t \eta\right)+A \wedge^\t \left( \eta \wedge \eta\right)\cr
&\,&+\,\d\eta \wedge^\t \b +\d\eta \wedge^\t d \eta +\d\eta \wedge^\t \left(A \wedge^\t \eta\right)\cr
&\,&+\,\d\eta \wedge^\t \left( \eta \wedge \eta\right)\,.\label{gft}
\end{eqnarray}
By using the following identities
$$ d(\eta\wedge\eta)+ \d\eta\wedge^\t d \eta = 0$$
$$ d(A\wedge^\t \eta) = dA \wedge^\t \eta - A\wedge^\t d\eta\,,$$
$$ A\wedge^\t (A \wedge^\t \eta) = (A \wedge A) \wedge^\t \eta \,,$$
$$ \d\eta \wedge^\t \b = -\d\b \wedge^\t \eta \,,$$
$$ A \wedge^\t (\eta\wedge \eta) + \d\eta \wedge^\t (A  \wedge^\t \eta )= 0\,,$$
$$ \d\eta\wedge^\t (\eta\wedge\eta) = 0 \,,$$
which follow from (\ref{patr}), (\ref{trpa}) and (\ref{jac}), the transformation (\ref{gft}) becomes
\begin{equation}
\G_{({A},\b)} \mapsto \G_{({A},\b)}+\H_{({{A}},\b)} \wedge^\t \eta \,.
\label{gt}\end{equation}

\subsection{BFCG action}

Given the 2-group gauge fields $(A,\b)$, one would like to find an action invariant under the gauge transformations (\ref{thin}) and (\ref{fat}), such that the corresponding equations of motion {imply the vanishing of the fake and the 2-curvature tensors, i.e.}
\begin{equation}\H_{(A,\b)}= \F_A - \d\b = 0 \,,\quad \G_{({A},\b)}= d\b + A\wedge^\t \b =0 \,.\label{2bf}\end{equation}

{The simplest way to obtain such an action is to consider the equations (\ref{2bf}) as the dynamical constraints and therefore to enforce them by using the Lagrange multiplier terms.} Since this was also the way of obtaining the action for a BF theory, where $\F_A =0$, we will then obtain a generalization of the BF-theory action. Therefore consider
\begin{equation}\label{S1} S_1 =   \int_M \langle B \wedge \H_{(A,\b)}\rangle_\lg +\int_M \langle C \wedge \G_{(A,\b)} \rangle_\le \,,\end{equation}
{where $\langle,\rangle_\lg$ and $\langle,\rangle_\le$ are $G$-invariant, {bilinear} non-degenerate and symmetric forms in the differential crossed module $(\le \to \lg,\t)$, see Definition 5. The} Lagrange multiplier field $B$ is a $\lg$-valued two-form, while the  Lagrange multiplier field $C$ is a $\le$-valued one-form. We will refer to this action as the BFCG action.

The action $S_1$ will be invariant under the thin gauge transformations if
\begin{equation} C \to \f^{-1}\t C \,,\quad B \to \f^{-1} B \f \,.\label{gbct}\end{equation}
This is ensured by the $G$-invariance of the bilinear forms $\langle ,\rangle_\lg$ and $\langle, \rangle_\h$, see Definition \ref{invariant}.

In order to make the action (\ref{S1}) invariant under the fat gauge transformations, the fields $B$ and $C$ have to transform as
\begin{equation}B \mapsto B+C\wedge^{\Tc} \eta \,,\quad C \mapsto C   \,,\label{bcf}\end{equation}
where  the antisymmetric map $\Tc\colon \le \times \le \to \lg$ is defined in \ref{ndg}. Note that $C\wedge^{\Tc} \eta$ is the antisymmetrization of $\Tc(C,\eta)$.

The BFCG action $S_1$ reduces to the BFCG action found in \cite{GPP} in the special case of a Lie crossed module of the form $\left(\d\colon \lg \stackrel{v \mapsto 0}{\mapsto} G,\t\right)$, where $\t$ denotes the adjoint action of $G$ on $\lg$, {and the abelian Lie group structure on $\lg$ is given by the sum of vectors. }

\begin{Proposition}
The BFCG action (\ref{S1}) is invariant under the gauge transformations (\ref{thin}), (\ref{fat}), (\ref{gbct}) and (\ref{bcf}). The equations of motion are given by
\begin{align}
 \H_{(A,\b)} =0 \,,\quad \G_{(A,\b)}&=0\,,\\ 
\d^*(B) + d C+A\wedge^\t C&=0 \,,\label{be}\\ 
d B+A\wedge B+\beta \wedge^{\Tc} C&=0 \,.\label{ae}
\end{align}
These equations of motion are obtained by calculating the variational derivatives of $S_1$ with respect to the fields appearing in it. Here $\d^*\colon \lg \to \h$ is obtained from the adjoint $\d^\dagger \colon \lg^* \to \h^*$, by using the isomorphisms $\lg^*\cong \lg$ and $\h^*\cong \h^*$ provided by the non-degenerate bilinear forms in $ \lg$ and $\h$. In components
$$ \d^* (B) = B^m \d^* (T_m) = B^m \d_m^{*\mu}  \tau_\mu  \,, $$
where $\d_m^{*\mu} = Q_{mn} \d^n_\nu q^{\nu\mu}$ and $q^{\nu\mu}=q^{\mu\nu}$ is the inverse matrix of $q_{\mu\nu}$.
\end{Proposition}

\subsection{Extended BFCG action}

Another way to obtain a 2-BF action is to introduce an additional auxiliary field 
$\a \in \A^1(M,\le)$ beside the Lagrange multiplier fields, such that
\begin{equation}S_2 =\int_M \langle B' \wedge \H_{(A,\b)}\rangle_\lg + \int_{M} \langle C \wedge \left( \G_{({A},\b)} + \H_{({A},\b)} \wedge^\t \a \right)\rangle_\le  \,.\label{ebfcg}\end{equation}
We have written $B'$ instead of $B$ because $B'$ will be invariant under the fat gauge transformations. 

It is easy to see that the action $S_2$ will be invariant under the thin gauge transformations if
\begin{equation}\a \to \f^{-1} \t \a \,, \quad B' \mapsto \f^{-1}B'\f \,,\quad C \mapsto \f^{-1} \t C \,,\end{equation}
while the invariance under the fat gauge transformations can be achieved if
\begin{equation} \a \to \a-\eta\,, \quad B' \mapsto B' \,,\quad C \mapsto C \,.\end{equation}

Note that both terms of the $S_2$ action are invariant under the gauge transformations, which does not happen in the case of the action $S_1$. As before, the invariance of the $S_2$ action under the thin gauge transformations is ensured by the $G$-invariance of the bilinear forms $\langle,\rangle_\lg$ and $\langle,\rangle_\h$, and by the identity (\ref{eqforinvariance}). Because $S_2$ contains an additional auxiliary field $\a$, we will refer to $S_2$ as an extended BFCG action.

It is not difficult to see that the actions $S_1$ and $S_2$  are related by the following field redefinition 
\begin{equation} B = B'-C \wedge^{\Tc} \a \,. \label{bbp}\end{equation}
The transformation 
$$(A,\b,C,B',\a) \stackrel{\Phi}{\mapsto} (A,\b,C,B'-C \wedge^{\Tc} \a)$$ 
is not invertible because $S_2$ has more fields. However, it can be shown that the dynamics of the theory defined by the action $S_2$ is determined by the dynamics of the $S_1$ theory.

The equations of motion for the action $S_2$ can be obtained by calculating the variational derivatives with respect to the fields appearing in the action. The variation with respect to $B'$ and $C$ gives
\begin{equation} \H_{(A,\b)} = 0 \,,\quad \G_{(A,\b)} + \H_{(A,\b)} \wedge^\t \a = 0 \,, \label{stf}\end{equation}
respectively. The variation with respect to $\b$ and $A$ can be obtained by substituting
(\ref{bbp}) into the equations of motion (\ref{be}) and (\ref{ae}) for $S_1$. There will be one more equation, corresponding to the variation of $S_2$ with respect to $\a$, and this one is given by
\begin{equation} C \wedge^\t \H_{(A,\b)} = 0 \,. \label{sta}\end{equation}

The equations (\ref{stf}) and (\ref{sta}) do not determine $\a$ so that $\a$ is determined only by the equations (\ref{be}) and (\ref{ae}) where $B= B'- C\wedge^\Tc \a$. Therefore, given a solution $(A,\b,B,C)$ of the $S_1$ equations of motion, then $(A,\b,B+C\wedge^\Tc \a,C, \a)$ is a solution of the $S_2$ equations of motion, where the components of $\a$ are arbitrary functions on $M$.

Note that the extended BFCG action can be easily modified by introducing powers of $B'$ and $C$ fields, in the following way 
\begin{equation}\label{ds2}
 S_2' = S_2 +  \int_M \lambda_1\langle B' \wedge B' \rangle_\lg + \lambda_2 \langle \big ( B'\wedge^\t C\big) \wedge C \rangle_\h +
\lambda_3 \langle \big (C \wedge^\T C\big) \wedge \big (C \wedge^\T C\big) \rangle_\lg \,,
\end{equation}
where $\lambda_1, \lambda_2$ and $\lambda_3$ are arbitrary constants and we have written only the non-trivial terms.
This modified action is clearly  invariant under thin and fat gauge transformations. 

The action (\ref{ds2}) is a generalization of the cubic term action in the case of three-dimensional BF-theory, as well as a generalization of the quadratic term BF-action in the four-dimensional case, see \cite{CCFM}.

\section{Conclusions}

The BFCG action (\ref{S1}) defines the dynamics of fake-flat and 2-flat 2-connections $(A,\b)$, as well as the dynamics of the auxiliary fields $B$ and $C$.  One can now use the action (\ref{S1}) to quantize the corresponding 2-BF theory. 

{A perturbative quantization of a BFCG theory can be obtained by using  Feynman diagrams, in the same way as the perturbative quantization of a BF theory can be obtained by using the corresponding Feynman diagrams, see \cite{CR1,CR2}. In this way one can obtain perturbative invariants of knotted surfaces, by using {the (non-abelian)}  2-holonomies (Wilson surface observables) defined in \cite{FMP1,FMP2,SW2}.}

{Due to the fact that these observables are trivial for the case of 2-flat and fake-flat 2-connections and 2-spheres  embedded in $S^4$, \cite{FMP1,FMP2}, {and given that the BFCG path-integral gives a delta function on the space of fake-flat and 2-flat 2-connections, it is very likely that the expectation values of these Wilson surface observables will be trivial invariants for embedded spheres in $S^4$}. The three-parameter deformation of the extended BFCG action  (\ref{ds2}) is a good starting point for solving this problem. In this case an appropriate modification of the observables will be needed, and this may yield non-trivial perturbative invariants of knotted surfaces. This belief is based on the analogy with the 3-dimensional BF theory case where the corresponding knot invariants can be modified in such a way, yielding Witten-Reshetikhin-Turaev invariants of links, see \cite{CCFM,FMM}. {A perturbative expansion of such a set of extended BFCG invariants could give a categorified version of Vassiliev knot invariants \cite{Ko,AF}.} 

{For the case of 4-manifolds without Wilson surfaces, a non-perturbative quantization of a BFCG theory can be obtained by using a generalization of the spin foam quantization method, see \cite{GPP}.}
Given a closed 4-manifold $M$ and a {Lie crossed module} $(\d : H \to G , \t )$, {the path-integral
$$Z(M)=\int \D A\,\D\b\, \D B\, \D C\, e^{i \int_{M} \langle B \wedge \F_{(A,\b)} \rangle_\lg +  \langle C \wedge \G_{(A,\b)} \rangle_\le }\,, $$
when properly defined, should coincide} with the volume of the space of fake-flat and 2-flat 2-connections over $M$, up to gauge transformations. {The $Z(M)$ could be defined by discretizing $M$ and by using the associated one and two-dimensional holonomies. This was the approach used in \cite{GPP}, where a  formal discretized expression $Z_{GPP}(M)$ for $Z(M)$ was given. In the case of a finite crossed module, $Z_{GPP}(M)$ coincides with Yetter's manifold invariant $Y(M)$, see \cite{Y,P1}. The} invariant $Y(M)$ was {originally} defined for three-dimensional closed manifolds, but the expression for $Y(M)$ can be proven to be an invariant of closed manifolds of arbitrary dimension \cite{P2,FMPo}. In fact,  $Y(M)$ depends only on the homotopy type of a manifold, and can be extended to general CW-complexes \cite{FMPo,FM1,FM2}.

Although $Y(M)$ is a homotopy invariant for a finite crossed module,  there is no immediate reason to believe that  $Y(M)$ will be a homotopy invariant for a general Lie crossed module.} It is an important open problem how to define $Y(M)$ in this case, and a possible approach is to generalize the construction of the Ponzano-Regge model path integral done in \cite{BNG} to the case of a Lie crossed module. Such an approach will require an extension of the Peter-Weyl Theorem to categorical representations of crossed modules, see \cite{E,BM}, {which appears not to have been addressed} {in the literature.

We expect that it will be possible to obtain a quantum group invariant related with $Z(M)$ by making a quantum-group regularization of the dual complex discretized expression of {$Z_{GPP}(M)$}. This expectation comes from the analogy with the BF-theory case, where the quantum group regularization of the dual complex state-sum representation of $Z(M)$ gives the Turaev-Viro invariant in the 3-manifold case, while in the 4-manifold case it gives the Crane-Yetter invariant. {Such an approach} will require a generalization of Lie crossed modules for the case of quantum groups, i.e. a quantum 2-group, see \cite{KL}, as well as a representation theory of quantum 2-groups at a root of unity. This quantum 2-group should be relevant for the quantization of the modified extended BFCG action (\ref{ds2}). 

Note that the construction of Lie crossed modules in Sect. 2 by using {chain complex}es of vector spaces provides {a lot of} non-trivial examples of Lie crossed modules where the group $H$ is non-abelian and the morphism $\d : H \to G$ is non-trivial, {such that the group $G$ is compact and non-abelian}. It also provides a matrix representation of Lie crossed modules, which is important for classical and quantum field theory considerations.

\bigskip
\noindent{\textbf{Acknowledgements}}

We would like to thank T. Strobl for discussions.

\end{document}